\documentclass[slac_one]{revtex4}
\usepackage{graphicx}
\usepackage{fancyhdr}
\def\met{\mbox{${\hbox{$E$\kern-0.6em\lower-.1ex\hbox{/}}}_T$}} %missing ET
\def\mex{\mbox{${\hbox{$E$\kern-0.6em\lower-.1ex\hbox{/}}}_x$}} %missing Ex
\def\mey{\mbox{${\hbox{$E$\kern-0.6em\lower-.1ex\hbox{/}}}_y$}} %missing Ey
\def\mexy{\mbox{${\hbox{$E$\kern-0.6em\lower-.1ex\hbox{/}}}_{x,y}$}} %missing Exy
 
\begin{document}

\title{Search for High Mass Standard Model Higgs at the Tevatron} %% Paper title goes here


\author{Michael Kirby (for the CDF and D\O\ collaboration)}
\affiliation{Northwestern University, Evanston, IL 60208, USA}
%
%\author{P. Lucas}
%\affiliation{FNAL, Batavia, IL 60510, USA}

\begin{abstract}
We report on searches for a standard model (SM) Higgs boson in $p\bar(p)$
collisions at a center of mass energy $\sqrt{s}=1.96$ TeV with the CDF and D\O\
detectors using an integrated luminosity of more than 3.0 $fb^{-1}$.  For a SM Higgs with mass greater than $135$ GeV$/c^2$, the dominant decay mode is two W bosons and the searches presented are based upon the subsequent electron and muon decays of the two W bosons.  Significant improvement in background modeling and signal predictions have been implemented since previous preliminary results.  No significant excess is observed, and limits on standard model Higgs production are calculated.  The observed 95\% confidence level upper limits are found to be a factor of 1.63 (2.0) higher than the predicted SM cross section at $m_H = 165$ GeV$/c^2$ for the CDF (D\O\ ) experiment while the expected limits are a factor of 1.66 (1.9) higher than the predicted SM cross section.
\end{abstract}

\maketitle

\thispagestyle{fancy}


\section{INTRODUCTION} % Section title should be in all capitals.

The standard model of particles physics incorporates a Higgs field in order to provide a mechanism for spontaneous symmetry breaking in the electroweak sector\cite{StandardModel}.  The SM though provides no prediction for the mass of the Higgs boson that is a consequence of this field, and to date no experimental data show evidence of a Higgs signal.  Current electroweak precision fits prefer a Higgs mass of less than 154 GeV$/c^2$ \cite{electroweak_fit} at 95\% confidence level.  Direct searches at the LEP-II experiments provide a lower limit  to the Higgs mass of 114.4 GeV$/c^2$\cite{lep_higgs_result} at 95\% confidence level.  At the Tevatron, the dominant production mechanism for a high mass ($m_H> 135$ GeV$/c^2$) Higgs is gluon fusion with the dominant decay mode of the Higgs to two W bosons.  For $m_H$ = 160 GeV$/c^2$, the branching ratio to two Ws is greater than 90\%.  In order to enhance signal to background significance, only the W decay modes including electrons and muons are used for the searches presented in these proceedings.  As well, the spin-correlations between the Higgs and the two Ws combined with the V-A decay of the W allows for kinematic discrimination between signal and background events based upon lepton opening angle.  Recent improvements in theory and calculations techniques have shown significant contributions from both vector boson fusion(VBF) and associated production with vector bosons.  These additional signal contributions and calculations have now been incorporated into high mass Higgs searches.  The CDF and D\O\  detectors are both general purpose particle detectors and have been described in detail elsewhere~\cite{CDF,D0} and will not be discussed in these proceedings.

\section{D\O\ $H\to WW(^*) \to \ell\ell '\nu\nu (\ell = e,\mu,\tau)$ SEARCH}

The $H\to WW(^*) \to \ell\ell '\nu\nu$ candidates are triggered using either a single lepton\footnote{In these proceedings, when referring to reconstructed leptons, only electrons and muons are considered.} or di-lepton trigger and corresponds to a integrated luminosity of 3.0 $fb^{-1}$.  Candidate events are selected by requiring large missing transverse momentum (\met) and either two electrons ($e^+e^-$), an electron and a muon ($e\mu$), or two muons ($\mu ^+\mu^-$).  Muons are reconstructed based upon hits in the muon chambers matched to reconstructed tracks and are required to have transverse momentum ($p_T$) greater than 10 GeV$/c$ and detector $|\eta|$<2.0.  Electrons are reconstructed by matching tracks with electromagnetic clusters, and are required to have $p_T$ greater than 15 GeV$/c$ and a detector $|\eta|<3.0$.  Both muons and electrons are required to be isolated from additional activity in the calorimeter and tracker.  The invariant mass of the two leptons in the event must be greater than 15 GeV$/c^2$.  Jets are reconstructed using a cone algorithm with $R = \sqrt{(\Delta\phi)^2+(\Delta\eta)^2} = 0.5$ and required to have $p_T^j > 15$ GeV$/c^2$.

To remove the dominant $Z/\gamma^*$ background, events are required to have missing transverse energy $\met >$ 20 GeV.  To discriminate between \met ~generated from mis-measurement and escaping neutrino energy, the variable $\met^{scaled}$ is constructed as:

\begin{equation}\label{et_scaled}
\met^{scaled}=\frac{\met}{\sqrt{\sum_{jets}(\Delta E^{jets}\cdot \sin{\theta^{jet}}\cdot \cos{\Delta\phi(jet,\met)})^2}}
\end{equation}

where the jet transverse energy resolution is approximated by $\Delta E^{jet}\cdot \sin{\theta^{jet}}$. The opening angle $\Delta\phi(jet,\met)$ between the projected energy fluctuation and the missing transverse energy provides a measure of the contribution of the jet energy resolution.  Additionally, the minimum transverse mass ($M_T^{min}$) is the lesser of the transverse masses calculated for each lepton as
\begin{equation}
M_T(\ell,\met) = \sqrt{2p^\ell_T\met(1-\cos{\Delta\phi(\ell,\met)})}
\end{equation}
The additional requirements for candidate events based upon these variables are optimized for each channel to optimized to suppress the dominate background in each final state. % and are listed in Table \ref{tab:d0_cut_table}.
The signal contribution is simulated using both gluon-gluon fusion and vector boson fusion processes where the cross sections have been normalized to next-to-next-to-leading order calculations \cite{nnlo_1,nnlo_2,nnlo_3}.

%\begin{table}[t]
%\begin{center}
%\caption{Summary of event selection for the three final states after preselection.}
%\begin{tabular}{|l|c|c|c|}
%\hline \textbf{Final State} & $e\mu$ & $e^+e^-$ & $\mu^+\mu^-$\\\hline\hline
%\met (GeV)& $>20$& $>20$& $>20$\\\hline
%$\met^{scaled}$&$>7$&$>6$&$>5$\\\hline
%$M_T(\ell,\met)$(GeV)&$>20$&$>30$&$>20$\\\hline
%$\Delta\phi(\ell,\ell)$&$<2.0$&$<2.0$&$<2.5$\\\hline
%\end{tabular}
%\label{tab:d0_cut_table}
%\end{center}
%\end{table}

\begin{figure*}[h]
\centering
\includegraphics[width=70mm]{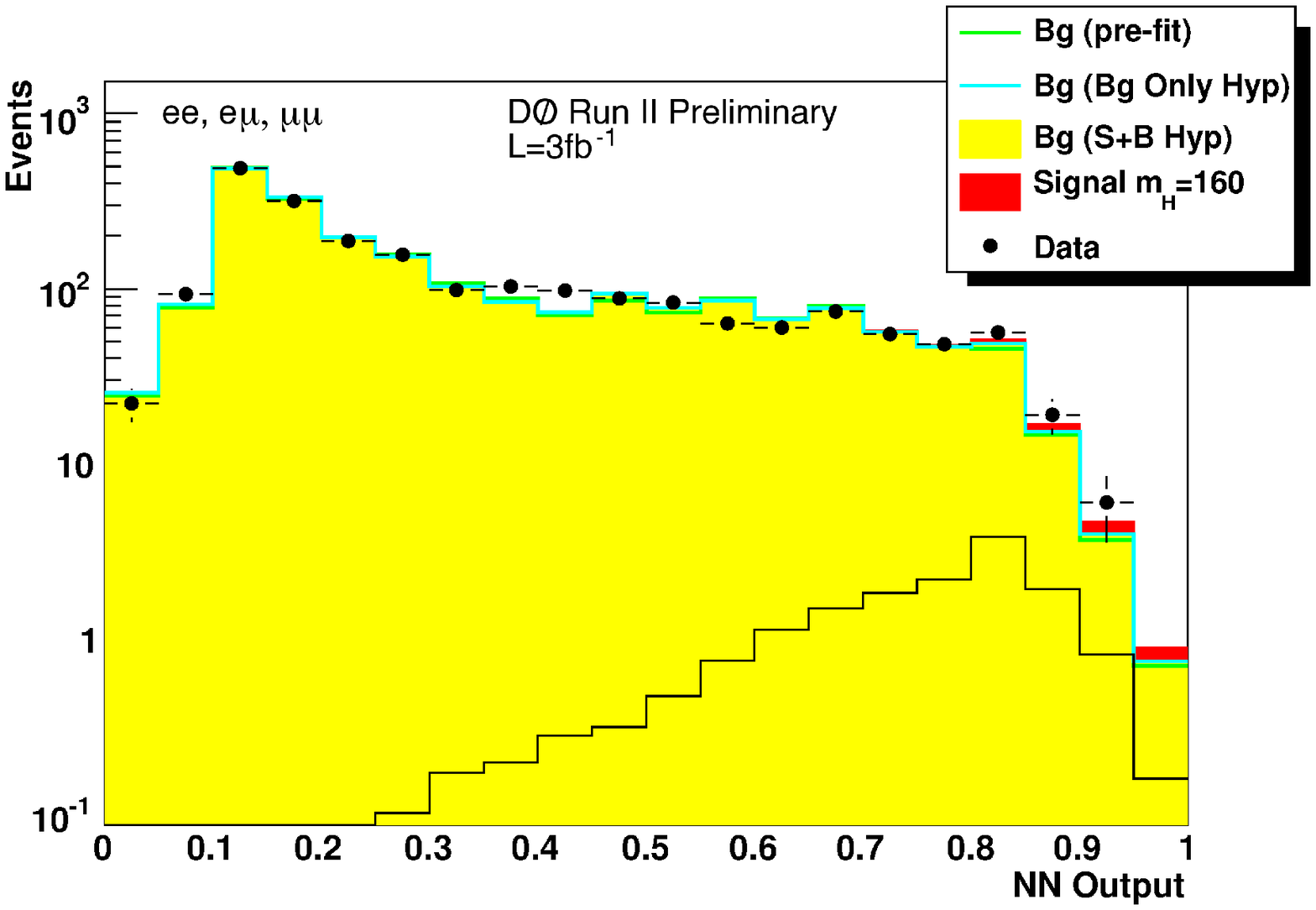}
\includegraphics[width=70mm]{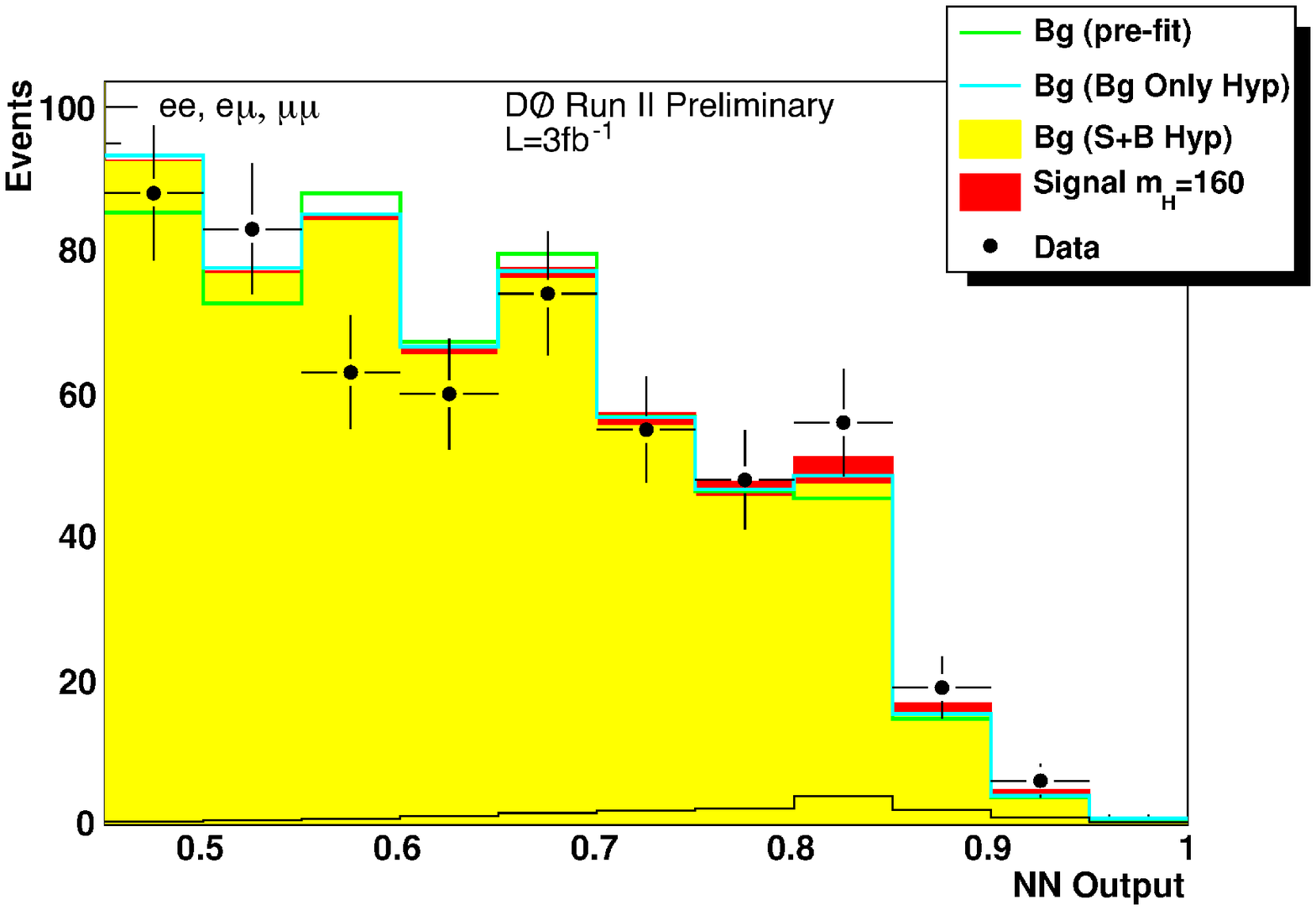}
\caption{Artificial Neural Net output after all event selections are applied for the combined D\O\ $e\mu$, $e^+e^-$, and $\mu^+\mu^-$ channels.  The output is shown on a logarithmic scale (left) and in a linear scale for the high NN output region (left).} \label{fig:d0_nnout}
\end{figure*}

\begin{figure*}[h]
\centering
\includegraphics[width=70mm]{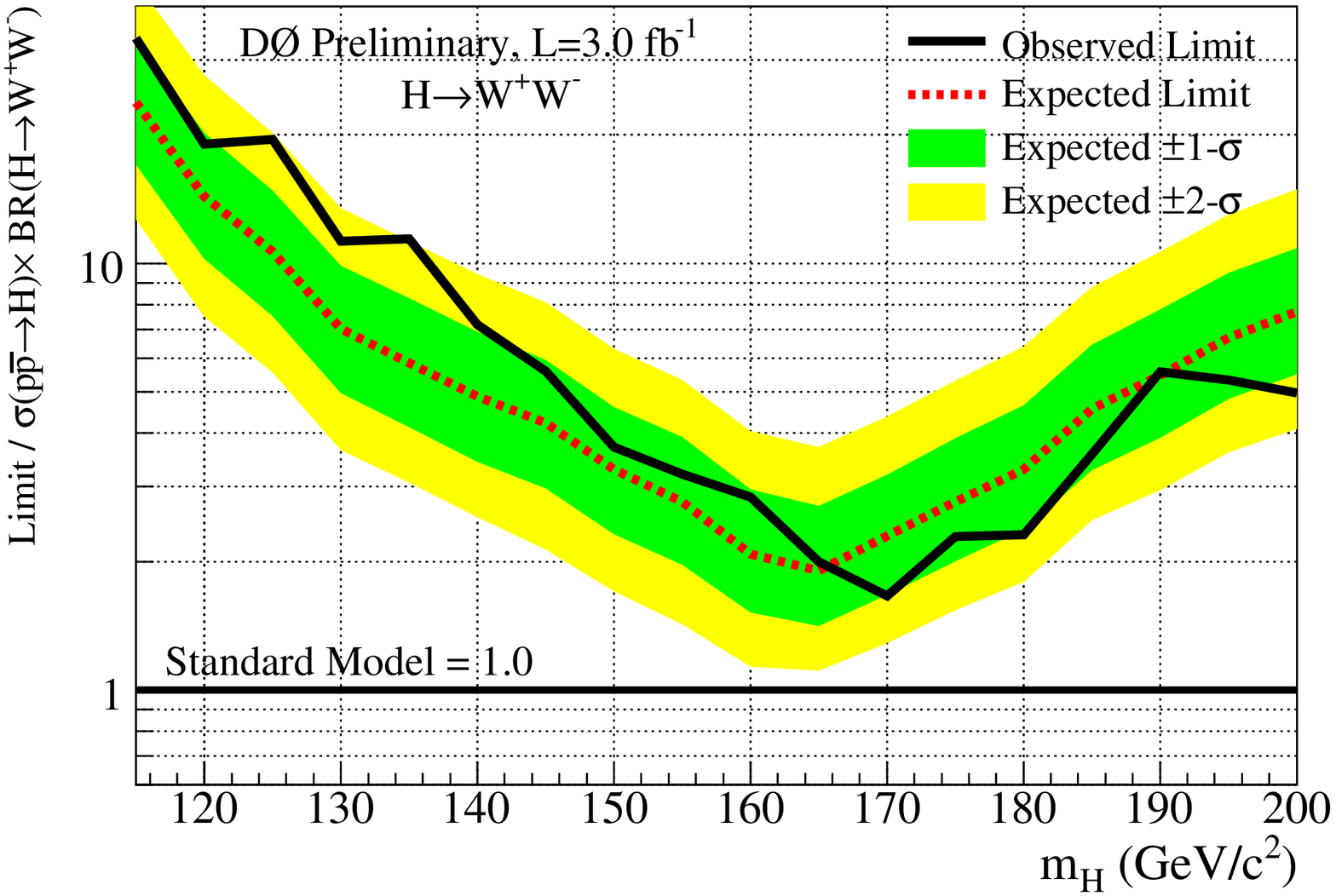}
\includegraphics[width=70mm]{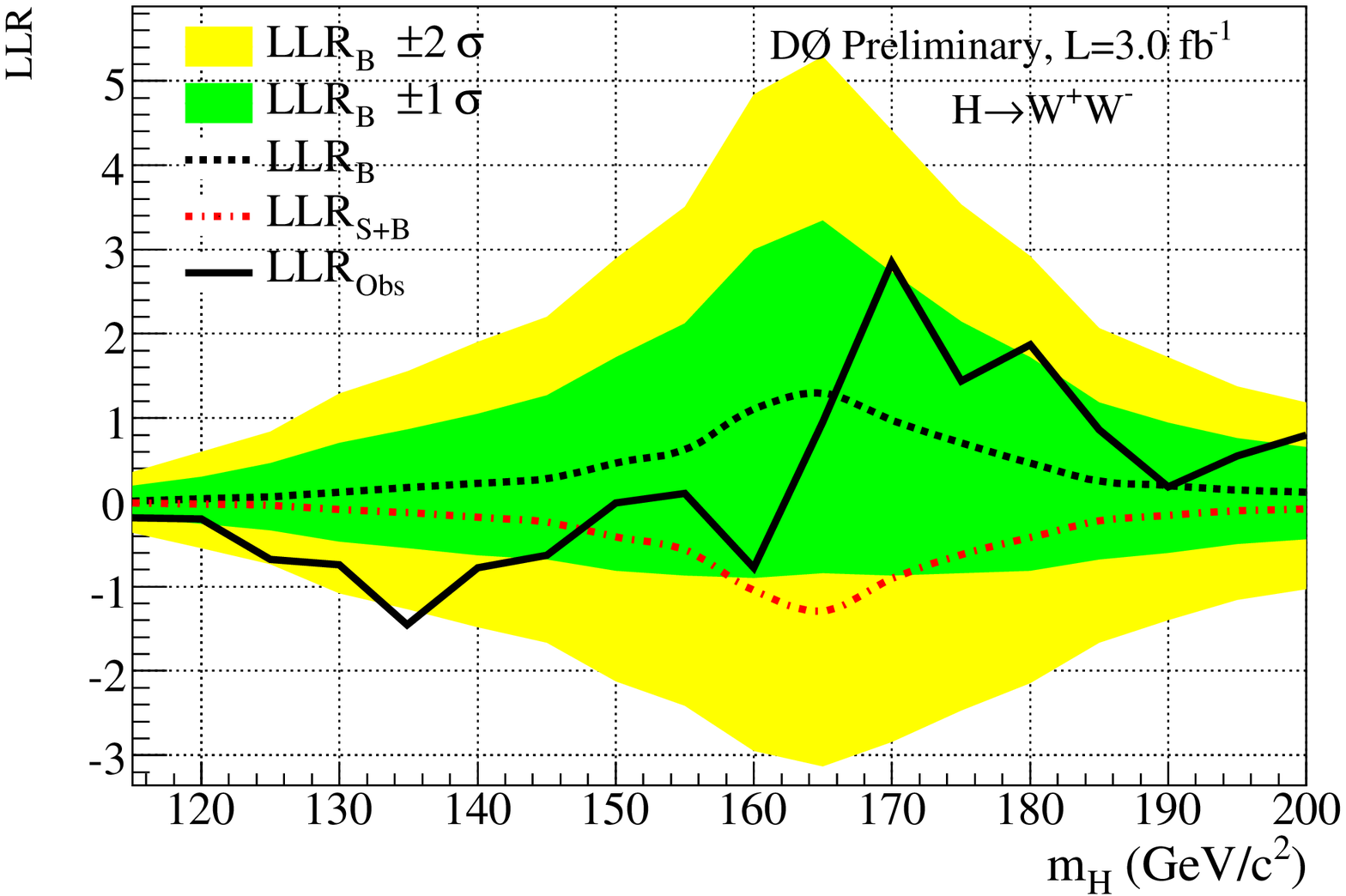}
\caption{The D\O\ calculated cross seciton limit for SM $H\to WW(*)$ production for Higgs masses between 115 and 200 GeV$/c^2$.  The left distribution show the log likelihood distribution of the combined $e\mu$, $e^+e^-$, and $\mu^+\mu^-$ channels.} \label{fig:d0_limits}
\end{figure*}

After all selection requirements have been applied, the combination of the three channels gives and expected 14.5$\pm$0.2 Higgs signal events for $m_H$ = 160 GeV$/c^2$ with a background expectation of $1714\pm 21$.  In order to increase sensitivity to a signal, an Artificial Neural Network (NN) is constructed using 11 input variables based upon reconstructed lepton kinematics, event kinematics, and event topology.  The NN output distributions for for the combined sample are shown in Figure ~\ref{fig:d0_nnout} and show no sign of an excess in the high NN output signal region.  Using a modified frequentist method that takes advantage of background dominated regions to minimize nuisance parameters, cross section limits on Higgs production for a range of masses are calculated from the log likelihood ratio distribution.  For $m_H=165$ GeV$/c^2$ the observed cross section limit was 2.0 times greater than the SM cross section with an expected sensitivity of 1.9 times the SM cross section.  The observed and expected cross section limits and log likelihood ratios for the full Higgs mass range are shown in Figure~\ref{fig:d0_limits}.

\vspace{1.0in}

\section{CDF $H\to WW(*) \to \ell\ell '\nu\nu (\ell = e,\mu,\tau)$ SEARCH}

Candidate events are triggered using either a central electron trigger, a central muon trigger, or a forward electron trigger that has an additional requirement on missing transverse energy in the calorimeter with the dataset corresponding to an integrated luminosity of 3.0$fb^{-1}$.  Leptons in the events are identified by requiring a central track matched to either an electromagnetic calorimeter cluster, a muon stub, or a non-fiducial region of the calorimeter and muon detector subsystems.  The leptons are categorized into one of six quality levels: two for electrons, three for muons, and one for non-fiducial tracks.  Events are then classified as contributing to either the high or low signal to background regions based upon the combination of the lepton quality categories.  All leptons are required to be isolated from other activity in the calorimeter and tracking subsystems.  Events are required to contain two lepton candidates with the first lepton having a minimum $E_T>20$ GeV and a looser requirement on the second leading lepton of $E_T>10$ GeV.  The lepton are required to have opposite charge and their invariant mass ($M_{\ell\ell}$)is required to be greater than 16 GeV$/c^2$.  To suppress the large Drell-Yan backgournd, the events are require to have $E_T^{spec} > 25$ GeV for dielectron and dimuon events, and $E_T^{spec}>15$ GeV for electron-muon events where $E_T^{spec}$ is defined as:

\begin{equation}
\met^{spec} \equiv \left \{ \begin{array}{ll} \met & if \Delta\phi(\met,lepton,jet) > \frac{\pi}{2}\\ \met\sin{(\Delta\phi(\met,lepton,jet))}& if \Delta\phi(\met,lepton,jet) < \frac{\pi}{2} \end{array} \right \}
\end{equation}

After the initial selection, the candidate events are analyzed based upon the jet multiplicity in the event.  A jet is required to have $p_T > 15$ GeV and $|\eta|<2.5$, and the three final states are the no jet events, 1-jet events, and events with 2 or more jets.  By separating the events based upon jet multiplicity, the contributions from the gluon-gluon fusion, VBF, and associate production can be partitioned and the appropriate multivariate technique applied.  The jet multiplicity was found to have excellent agreement with the predicted background giving confidence to analyzing the candidate events  in several jet samples.

\subsection{0-Jet Analysis}
For events with a jet veto, gluon-gluon fusion is the only signal production process considered to contribute.  Using MCFM\cite{mcfm} LO calculations, a likelihood ratio discriminant is calculated as a ratio between the the integrated probability density of $H\to WW$ or $WW$ production and the sum of the probability densities of $H\to WW$, $WW$, $ZZ$, $W\gamma$, and $W+jets$ production for each event.  Because of the jet veto, there is higher precision in the measured \met, and the integration over unobserved degrees of freedom provides more discriminating power.  The two likelihood ratios are then input into an Artificial Neural Network along three other variables: $\Delta R(\ell,\ell)$, $\Delta\phi (\ell,\ell)$, and the scalar sum of the leptons $E_T$ and \met.  The resulting NN output in the high signal to background sample for jet veto events is shown in Figure~\ref{fig:cdf_nnout}.

\subsection{1-Jet Analysis}
For events containing one jet, the signal contribution from gluon-gluon fusion is combined with the contributions from VBF and associated production \cite{pythia} adding an additional 20\% expected signal.   An Artificial Neural Network is constructed using eight input variables: $M(\ell,\ell)$, the transverse mass $M_T(\ell\ell\met)$,$\Delta R(\ell,\ell)$,$H_T$, $\met^{spec}$,$p_T(\ell_1)$,$p_T(\ell_2)$,$E(\ell_1)$.  The resulting NN output in the high signal to background sample for one jet events is shown in Figure~\ref{fig:cdf_nnout}.

\subsection{2-Jet Analysis}
\begin{figure*}[t]
\centering
\includegraphics[width=60mm]{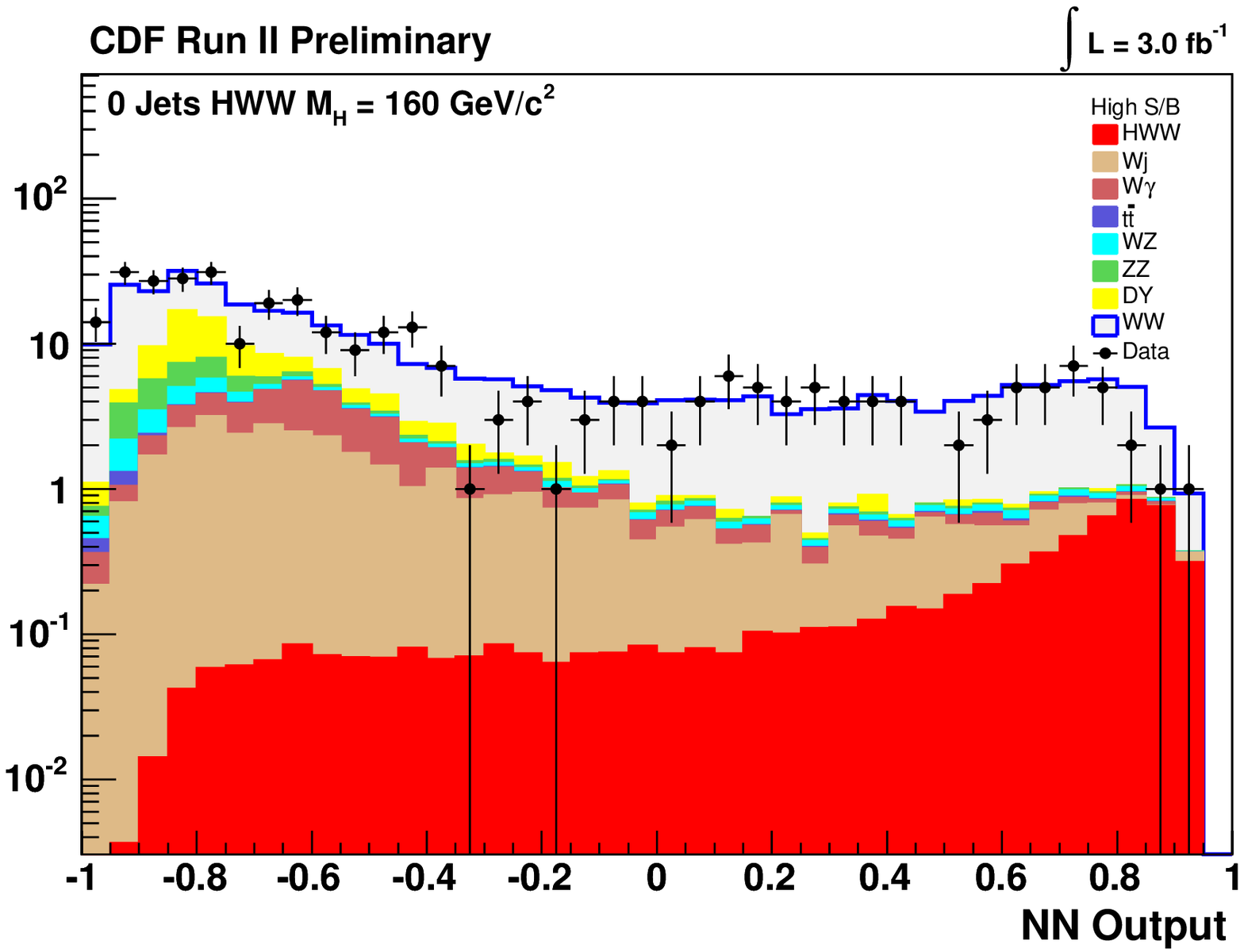}
\includegraphics[width=60mm]{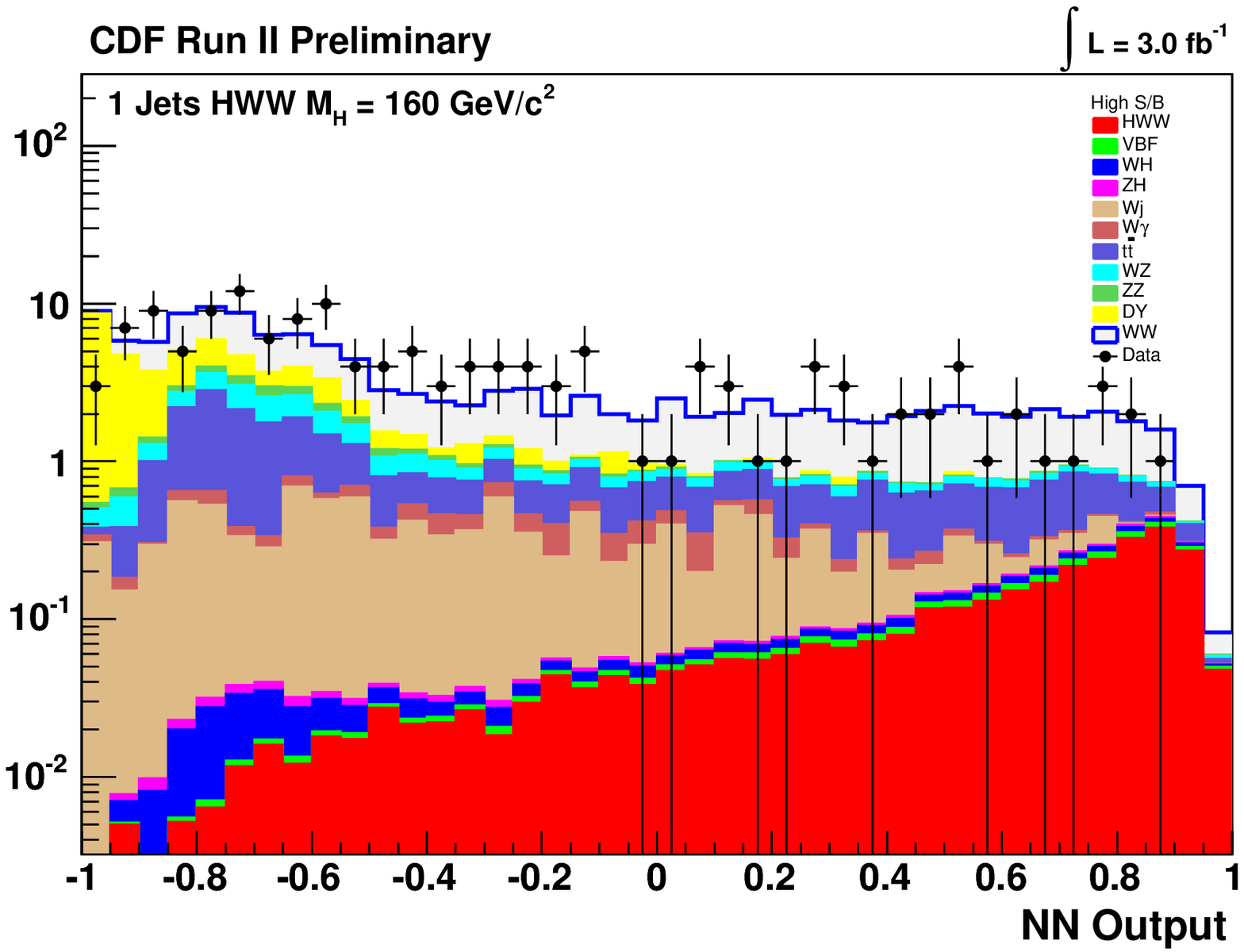}
\includegraphics[width=60mm]{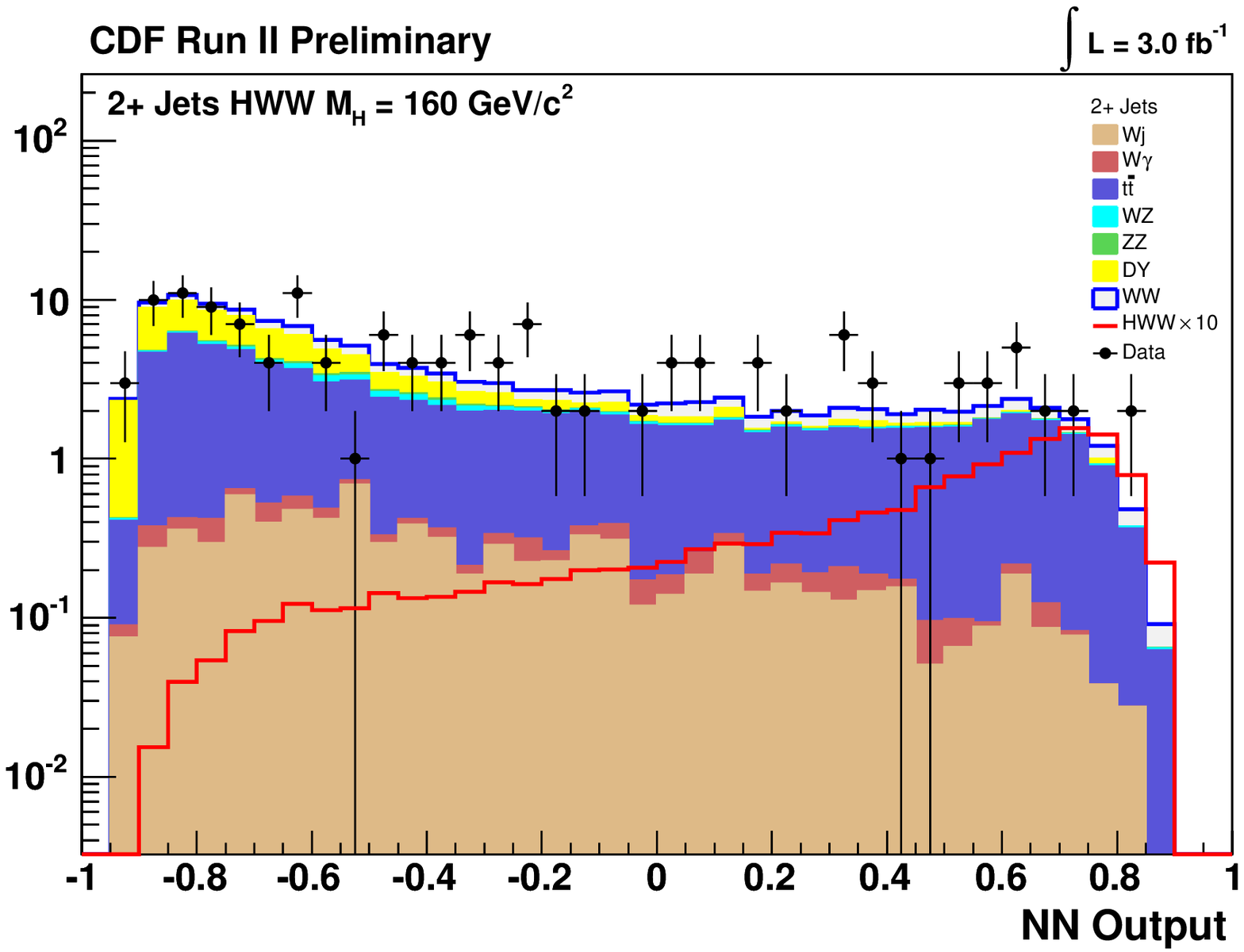}
\caption{Artificial Neural Net output after all event selections are applied for the 0-jet, 1-jet, and 2 or more jet channels for the CDF analysis.  The output is shown on a logarithmic scale for the high signal to background samples.} \label{fig:cdf_nnout}
\end{figure*}

For events containing two or more jets, the signal contribution from gluon-gluon fusion is combined with the contributions from VBF and associated production adding an additional 60\% expected signal.   An Artificial Neural Network is constructed using eight input variables: $M(\ell,\ell)$, $E_T(\ell_1)$, $E_T(\ell_2)$,$\Delta R(\ell,\ell)$,$H_T$, $\Delta\phi (\ell,\ell)$,$\Delta\phi (\vec{\ell} +\vec{\ell},\met)$, $\vec{p_T^{jet1}}+\vec{p_T^{jet2}}$.  The resulting NN output in the high signal to background sample for two or more jet events is shown in Figure~\ref{fig:cdf_nnout}.

\subsection{Combined CDF Results}

\begin{figure*}[t]
\centering
\includegraphics[width=60mm]{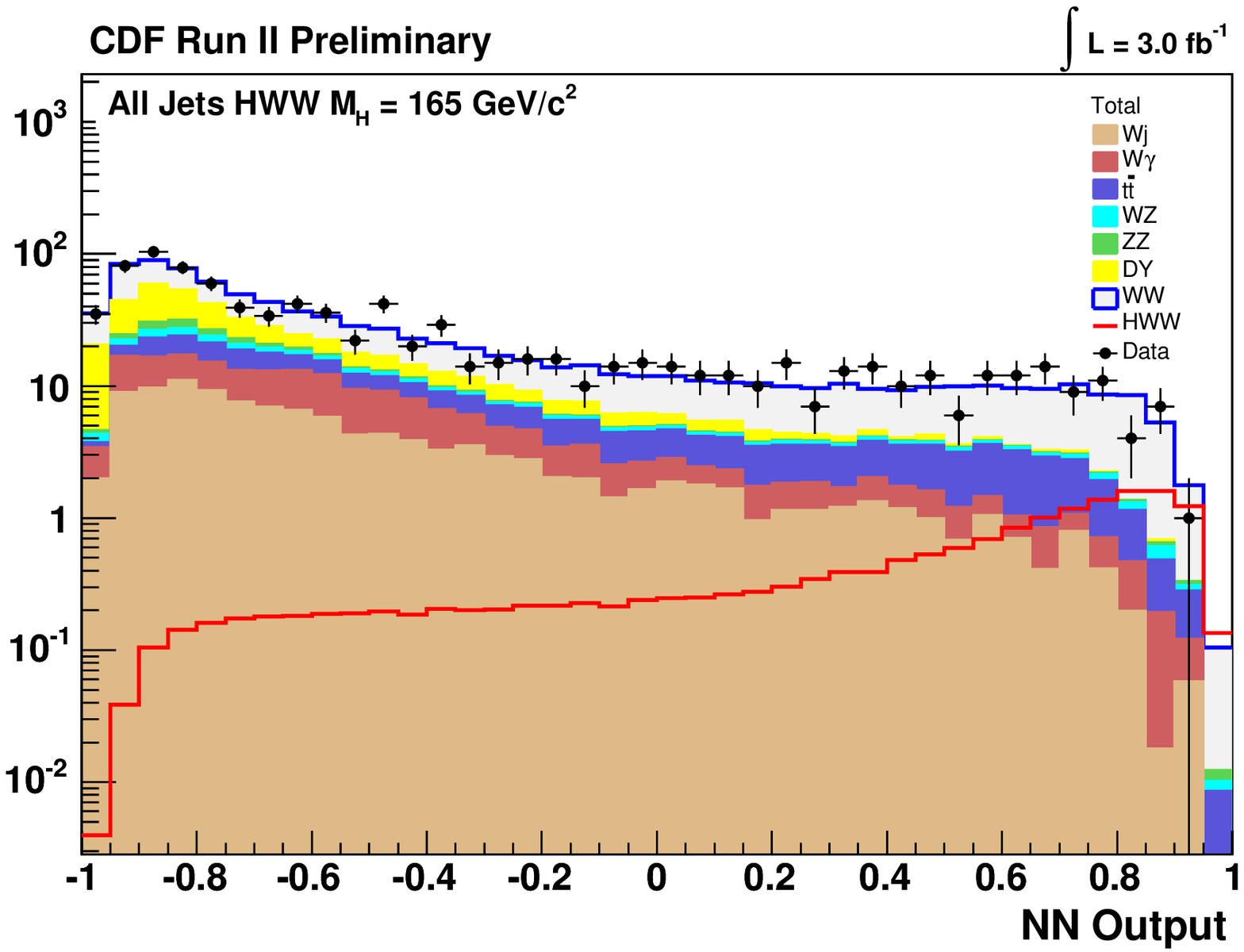}
\includegraphics[width=60mm]{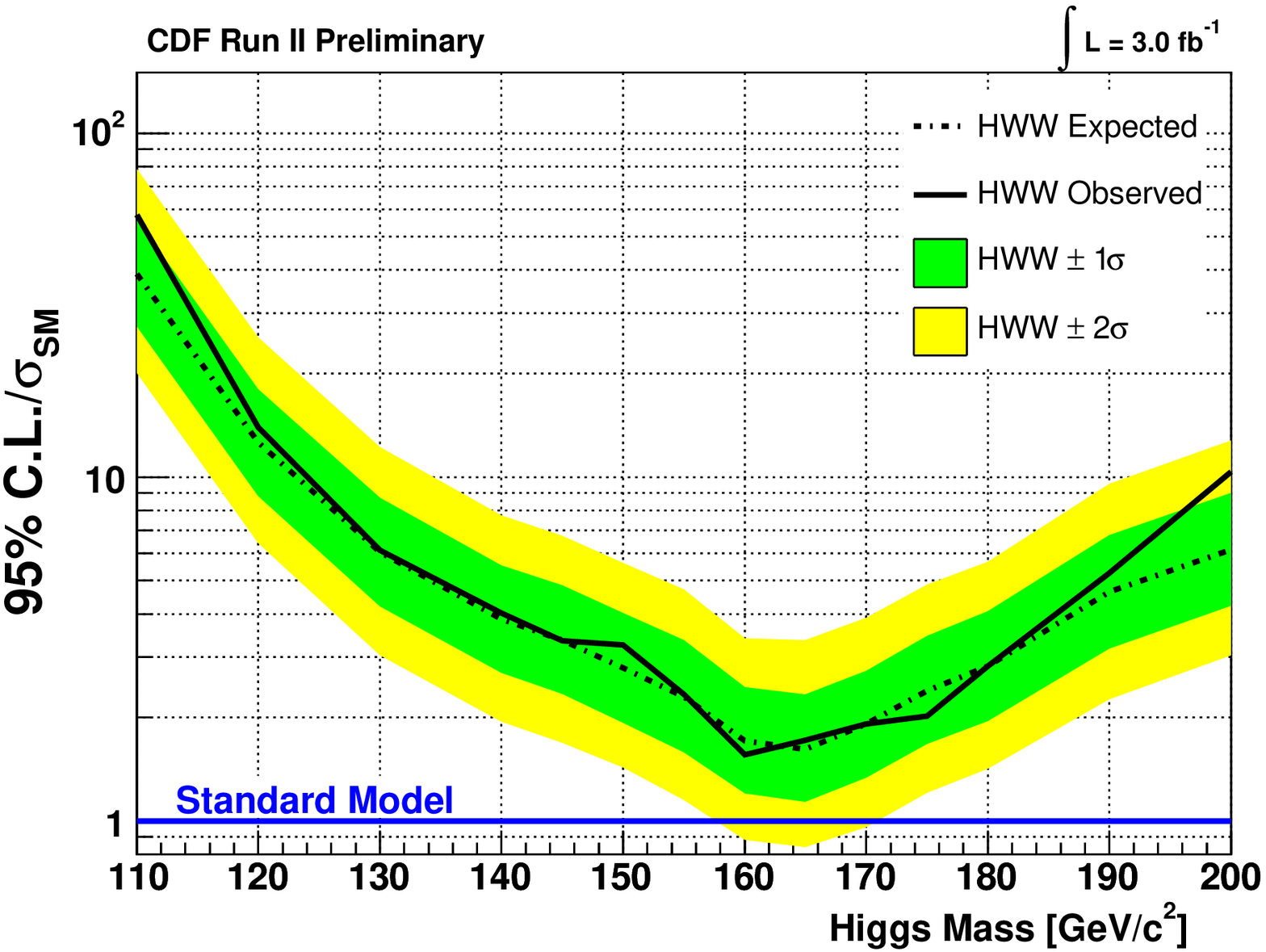}
\caption{Artificial Neural Net output after all event selections for the combined sample are shown in the left plot.  The CDF observed and expected cross section production limit as a ratio to the SM cross section are shown in the right plot.} \label{fig:cdf_limits}
\end{figure*}

The NN outputs for the five samples (two for 0-jet, two for 1-jet, and one for 2-jet) are combined into a binned likelihood function.  The signal to background rates are allowed to float in each bin based upon a set of Gaussian constraint while maintaining the expected signal to background ratio.  The 95\% confidence limit is determined based upon 10,000 Monte Carlo background only experiments generated from the expected yields keeping systematics between channels correctly correlated.  For $m_H=165$ GeV$/c^2$ the observed cross section limit was 1.63 times greater than the SM cross section with an expected sensitivity of 1.66 times the SM cross section.  The combined NN output for all channels and the observed and expected limits for the combined sample is shown in Figure~\ref{fig:cdf_limits}.

\section{CONCLUSION}
The search for standard model $H\to WW(*)$ production in the $\ell\ell\nu\nu$ final state was performed at CDF and D\O\.  The searches show no signs of an excess in the area of sensitivity and have placed limits of 1.63 and 1.9 times the SM cross section for CDF and D\O\ respectively.  The Higgs search analyses at the Tevatron continue to show factors of improved sensitivity beyond that expected from the simple enlargement of the data sample.  The CDF and D\O\ combined cross section limit for the $H\to WW(*)$ production was performed and provided the first 95\% exclusion of a standard model Higgs boson at the Tevatron at $m_H=170$ GeV$/c^2$.  The search for the Higgs will continue to be the focus of the Tevatron program and the sensitivity should improve considerably in the coming years.

\begin{acknowledgments}
The author wishes to thank the Fermilab Accelerator Division, the CDF and D\O\ Collaborations, and the faculty and staff of Northwestern University.  The author would also like to thank the organizers of the conference for an excellent program and wonderful experience.
Work supported by Department of Energy.
\end{acknowledgments}


\begin{thebibliography}{99}   % Use for  1-9  references
%\begin{thebibliography}{99} % Use for 10-99 references

\bibitem{StandardModel} P. W. Higgs, Phys. Rev. Lett. {\bf 13}, 508 (1964).
\bibitem{electroweak_fit} The LEP Electroweak Working Group, http://lepewwg.web.cern.ch/LEPEWWG/
\bibitem{lep_higgs_result} R. Barate et al., Phys. Lett. B {\bf 565}, 61 (2003).
\bibitem{CDF} The CDF Collaboration, FERMILAB-PUB-96-390-E.
\bibitem{D0}
%    Standard D\O\ detector reference:  \\
D\O\ Collaboration, V. Abazov {\it et al.}, ``The Upgraded D\O\ Detector,''
Nucl. Instrum. Methods Phys. Res. A {\bf 565}, 463 (2006).

\bibitem{nnlo_1}S. Catani, D. de Florian, M. Grazzini and P. Nason, JHEP {\bf 0307}, 028 (2003) [arXiv:hep-ph/0306211].
\bibitem{nnlo_2}K.A. Assamagan et al., [Higgs Working Group Collaboration], arXiv:hep-ph/0406152.
\bibitem{nnlo_3}U. Aglietti, R. Bonciani, G. Degrassi, A. Vicini, arXiv:hep-ph/0610033.
\bibitem{pythia}T. Sjostrand, S. Mrenna, and P. Skands, JHEP {\bf 05}, 026 (2006).
\bibitem{mcfm}J. Campbell and K. Ellis, MCFM - Monte Carlo for FeMtobarn processes, http://mcfm.fnal.gov/.

\end{thebibliography}
\end{document}